\documentclass[conference]{IEEEtran}
\usepackage{epsfig}
\usepackage{times}
\usepackage{float}
\usepackage{afterpage}
\usepackage{amsmath}
\usepackage{amstext}
\usepackage{amssymb,bm}
\usepackage{latexsym}
\usepackage{color}
\usepackage{graphicx}
\usepackage{amsmath}
\usepackage{amsthm}
\usepackage{graphicx}
\usepackage[center]{caption}
\usepackage{pstricks}
\usepackage{caption}
\usepackage{subcaption}
\usepackage{booktabs}
\usepackage{multicol}
\usepackage{lipsum}% dummy text
 \usepackage[T1]{fontenc}

\allowdisplaybreaks

\newtheorem{thm}{Theorem}%[section]

\newtheorem{lem}{Lemma}

\begin{document}

\title{On the Optimality of Uncoded Cache Placement}

\author{\IEEEauthorblockN{Kai~Wan}\IEEEauthorblockA{Laboratoire des Signaux et Système (L2S)\\
CentraleSupélec-CNRS-Université Paris-Sud\\
Gif-sur-Yvette, France\\
Email: kai.wan@u-psud.fr}\and \IEEEauthorblockN{Daniela~Tuninetti}\IEEEauthorblockA{University of Illinois at Chicago\\
Chicago, IL 60607, USA\\
Email: danielat@uic.edu}\and \IEEEauthorblockN{Pablo~Piantanida}\IEEEauthorblockA{Laboratoire des Signaux et Système (L2S)\\
CentraleSupélec-CNRS-Université Paris-Sud\\
Gif-sur-Yvette, France\\
Email: pablo.piantanida@centralesupelec.fr}}
\maketitle
\begin{abstract}
Caching is an efficient way to reduce peak-hour network traffic congestion
by storing some contents at user's local cache without knowledge of later
demands. Maddah-Ali and Niesen initiated a fundamental study of
caching systems; they proposed a scheme (with uncoded cache placement and linear network coding delivery) 
that is provably optimal to within a factor $12$. 
In this paper, by noticing that 
when the cache contents and the demands are fixed, the caching
problem can be seen as an \emph{Index Coding} problem, we show 
the optimality of Maddah-Ali and Niesen's scheme assuming that cache placement is restricted to be uncoded and
the number of users is not less than the number of files.
Furthermore, this result states that further improvement to the Maddah-Ali and Niesen's scheme in
this regimes can be obtained only by {\it coded cache placement}.
%By the help of the
%index coding outer bounds in \cite{onthecapacityindex}, we prove
%the optimal scheme of the considered problem under the constraint
%of uncoded cache placement is to firstly let each user have identical
%cache size and then use the coded caching scheme in \cite{dvbt2fundamental}.
%This result can also cover the caching problem with identical cache
%size. In other words, if we need a lower load coded cache placement
%is necessary.
\end{abstract}

\IEEEpeerreviewmaketitle{}

\section{Introduction}
\label{sec:intro}

%\subsection{Background of caching system and index coding}

%Nowadays 
Caching is %used in more and more cases 
a widely used technique to reduce traffic congestion
during the peak hours.
%%Since the request amount is variable depending on time, 
%The main idea %of caching 
%is to duplicate some content
%at the user's terminal called cache. 
%When a user requests a file in a peak hour, 
%he can make use of the information stored in
%his cache so that the network load can be reduced. 
%%So a whole process with 
Caching can be divided into two phases:
%In the first phase called
placement phase (each user stores some contents in his cache during
the off peak hours without knowledge of later requests) 
%In the second phase called 
and delivery phase (after receiving the connected users' requests and according to their cache contents the central server broadcasts
packets to users).

%A popular caching network system is introduced 
A fundamental study of caching systems appears in \cite{dvbt2fundamental}:
%A center server equipped with 
where a server has $N$ identical-length files and is connected
to $K$ users. In the placement phase, users store pieces of files within
their cache of size $M$. 
%cooperatively. The cooperation can be done
%because in a centralized caching system these $K$ users are connected
%to the server during the two aforementioned phases. 
In the delivery
phase, each user demands one specific file from the server. 
Based on the users' demands and cache contents,
packets are broadcasted over an error-free
shared link from the server to all the users. 
The objective is 
%with the fixed memory size to find the minimum 
minimize the number of transmission, or 
load, in the delivery phase for
the worst-case demands.
% among all the possible demands. In \cite{dvbt2fundamental},
%an efficient caching approach is proposed, called 
The {\it coded caching} approach of \cite{dvbt2fundamental}
uses combinatorial cache construction in the placement phase
and linear network coding in the delivery phase;
the scheme is shown to achieve an additional
{\it global caching gain} compared to the conventional local caching
gain of uncoded systems.
%Note that this 
The method in \cite{dvbt2fundamental} uses uncoded cache placement, 
%i.e., each file is divided into several disjoint parts and each user directly stores
%one or several parts in his cache. 
but examples are given to show that coded cache placement performs better in general.
By using a cut-set 
%information theoretic
outer bound for the min-max load,
the scheme of \cite{dvbt2fundamental} is shown to be optimal to
%that its achievable load is 
within a factor of $12$.
%of the cut-set outer bound.

%Several works are done for the same problem in \cite{dvbt2fundamental}.
In \cite{criticaldatabase}, \cite{improvedlower}, \cite{noteonfundamental}
and \cite{ISIT2015outerbound}, outer bounds tighter than the cut-set bound provided in \cite{dvbt2fundamental} were proposed. An improved inner bound was proposed in \cite{groupcastindexcoding},
whose achievable load is equal to the fractional local chromatic number
(described in \cite{indexcodingrandom}) of a directed graph formed
by the users' demands and the cache contents. In \cite{smallbufferusers}, 
it is shown that when $N\leq K$ and $MK\leq1$ %\frac{1}{K}
(i.e., small cache size regime)
a scheme based on coded cache placement achieves the cut-set outer bound
and it is thus optimal.

Variations on the basic model of \cite{dvbt2fundamental} so as to
account for decentralized cache placement, non-uniform cache sizes,
non-uniform demands, non-uniform file-sizes, etc., have been
considered in the literature, but we do not 
summarize them here for sake of space.
In general, the question of the exact optimality of the achievable 
scheme proposed in \cite{dvbt2fundamental} is open.
This work makes progress into this direction.

{\bf Contributions.}
Our main result is based on the following observation.
When the users' demands and cache contents are fixed, the delivery phase can
be seen as an index coding problem \cite{indexcodingrandom},
\cite{indexcodingwithsi} and \cite{onthecapacityindex}. 
%So transforming
%a cache problem with fixed demands and caches to an index coding problem
%is a doable approach to solve the the caching problem. 
For the index coding problem,
an outer bound based on the sub-modularity of entropy
is proposed in \cite[Theorem 1]{onthecapacityindex}
and loosened in \cite[Corollary 1]{onthecapacityindex}.
%,  and its relaxed version . 
Although %the outer bound in 
\cite[Corollary 1]{onthecapacityindex}
is not generally tight, it is fairly simple and thus often used.
%it can always give us the optimal capacity region. 
%Therefore, this outer bound is used in many cases. For example,
%it is used for solving the linear topological interference schemes
%in \cite{toplogical}.
%
We exploit  \cite[Corollary 1]{onthecapacityindex}
%Our intention is to show how to utilize index coding outer bounds
%in \cite{onthecapacityindex} into a  cache problem similar to the
to derive a `converse' for the scheme in \cite{dvbt2fundamental};
we actually further relax the original setting in \cite{dvbt2fundamental}
%is that instead of assuming all the users have the same cache size our  considered problem 
by considering a constraint on the sum of the cache size of all users and on the sum of the total length of all files, in contrast  to assuming that each cache size is equal and each file length is equal.

%By the help of a linear programming based on in \cite[Corollary 1]{onthecapacityindex},
Our main result shows that under the constraint of 
uncoded cache placement and $N\geq K$,
%the solution to achieve 
the minimal load of the worst case among all the possible demands is achieved by %letting each user have the same cache size,  each file have the same size, and using 
the two-phase strategy in \cite{dvbt2fundamental}, even when the system is relaxed so as to allow optimal cache size allocation among users, subject to a sum cache size constraint,
and optimal file size allocation, subject to a sum file size constraint.

%Moreover, our proof shows the optimality of the two-phase strategy in \cite{dvbt2fundamental} under the constraint of uncoded cache placement for the centralized caching system with identical cache size and identical file length. 
It is worth to mention that past work on caching has mainly focused on tightening the outer bound of \cite{dvbt2fundamental} for $N\geq K$ rather than the inner bound. Our result shows that the inner bound \cite{dvbt2fundamental} can not be improved, unless coded cache placement is considered, as in \cite{smallbufferusers}.
%It is valuable to show it because while people doing the research on this problem have found some tighter outer bounds than \cite{dvbt2fundamental}, few have found a strictly better achievable bound when $N\geq K$ except \cite{smallbufferusers}. So the proof in this paper will give people an important fact that if we want to design a scheme with lower load, coded cache placement is necessary.
%
%This result can also cover the identical cache size problem. 
We also note that the inner and outer bound of \cite{dvbt2fundamental} coincide when $M\geq N(1-1/K)$ (i.e., large cache size regime).
%for the identical cache size and file length problem the general optimality of the coded caching scheme in \cite{dvbt2fundamental} if $M$ is large.
An interesting question that emerges from these results is where the optimal load $L$ vs per-user cache size $M$ is the same when the role of $L$ and $M$ are swapped.

{\bf Paper Outline.}
The remainder of the paper is organized as follows. 
Section~\ref{sec:model} presents the system model and past results from \cite{dvbt2fundamental}. %and past work. 
Section~\ref{sec:main} shows the main result of this paper. 
Conclusions and the further work are discussed in Section~\ref{sec:conc}.
Proof can be found in Appendix.

{\bf Notations.}
Calligraphic symbols %(e.g., $\mathcal{J}$) to 
denotes sets.
$|\cdot|$ is %used to represent 
the cardinality of a set or the length of a file. 
We denote $[1:K]:=\left\{ 1,2,\ldots,K\right\}$ and 
$\mathcal{A\backslash  B}:=\left\{ x\in\mathcal{A}|x\notin\mathcal{B}\right\} $.
$\oplus$ represents the bit-wise XOR operation (zeros may need to be appended
to make the vectors have the same length).
%are added in the end
%of the shorter term such that the two terms beside $\oplus$ have
%the same length.  $B(n,k)$ is 
The binomial coefficient is indicated as
$B(n,k):={n \choose k}$. The number of $k$-permutations of $n$ is indicated as $P(n,k):=n\cdot(n-1)\cdots(n-k+1)$.
%%\left(\begin{array}{c}
%%n\\
%%k
%%\end{array}\right)
%=\frac{n!}{k!(n-k)!}$, and the factorial is $n!$.

\section{System Model and Known Results}
\label{sec:model}

%In this part, we will 
We start by describing the considered system model and discussing
some existing results from \cite{dvbt2fundamental}.
%related to this problem. 
Then we introduce the necessary tools from index coding in \cite{onthecapacityindex}.
% problem is
%introduced in the end of this part.
%\subsection{System model}

\begin{figure}%[H]
\centering
\includegraphics[scale=0.5]{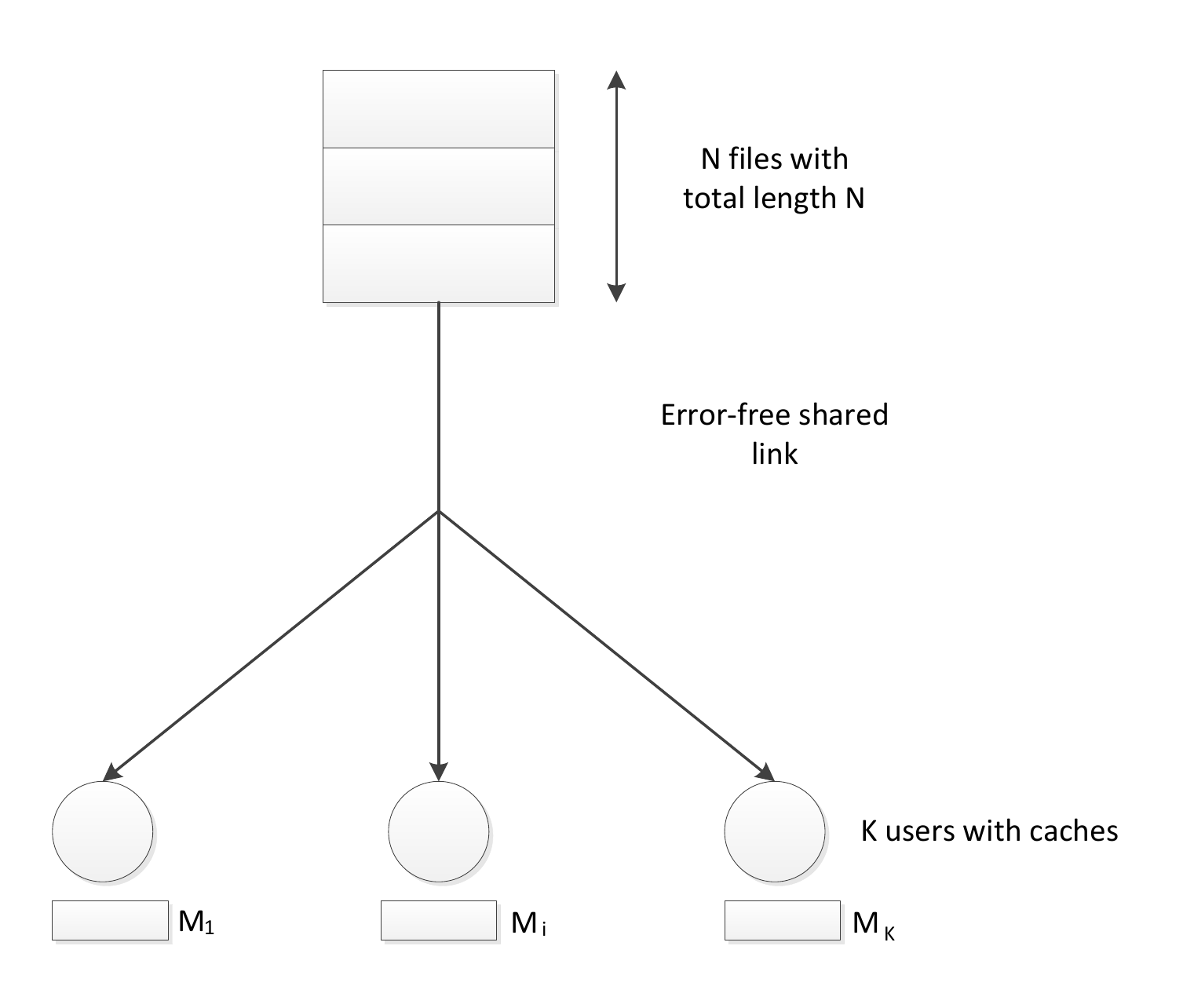}%\protect
\caption{Broadcast caching system.}
\label{fig:model}
\end{figure}

We consider the broadcast system model with caches illustrated in
Fig.~\ref{fig:model}.
A central server, equipped with $N$ files, $(F_{1},F_{2},\ldots,F_{N})$,
is connected to $K$ users through a broadcast link. 
There is a sum file size constraint $\frac{1}{N}\sum_{i\in[1:N]}|F_{i}|\geq 1$
(``$1$'' is one unit of file; load is measured in the same unit).
%In this paper, we consider a simple case where 

In the placement phase, user $i\in[1:K]$ stores information about the $N$ files in his cache
of size $M_{i}$, where $M_{i} \geq 0$. %\in[0,N]
This phase is done without knowledge of users' demands. 
There is a sum cache size constraint $\frac{1}{K}\sum_{i\in[1:K]}M_{i}\leq M$.

In the delivery phase, user $i\in[1:K]$ demands
one file $F_{d_{i}}$ where $d_{i}\in[1:N]$. Given the demand vector 
$\mathbf{d}=(d_{1},d_{2},\ldots,d_{K})$
the server transmits a signal of length $L$ through an error-free
shared link. Then, each user can restore his desired file by using the received
data through the shared link and the side information provided by his cache.

A memory-load $(M_{1},M_{2},\ldots,M_{K},L)$ is said to be achievable
for $(F_{d_{1}},F_{d_{2}},\ldots,F_{d_{K}})$ if every user can recover
his desired file with high probability. 
We want to minimize $L$ for the worst-case demands.

%subject to the {\it total size} $\sum_{i\in[1:K]}M_{i}\leq KM$, for some $M$ and $\sum_{i\in[1:N]}|F_{i}|\geq N$ 

%find the optimal memory-rate tradeoff $\sum_{\overset{K}{\sum_{i=1} }M_{i}=KM}{\mathrm{argmin}}L^{*}(M_{1},\ldots,M_{K})$, where $L^{*}(M_{1},\ldots,M_{K})=\mathrm{inf}\{L:(M_{1},\ldots,M_{K},L)\thinspace\thinspace\mathrm{is\thinspace\thinspace achievable\thinspace\thinspace for\thinspace\thinspace all\thinspace\thinspace(F_{d_{1}},\ldots,F_{d_{K}})}\}$.
%In other words we want to minimize load of the worst case under the
%constraint that the total cache size is fixed. 

%In general this problem is not solved so far. Instead several inner
%bounds and outer bounds are proposed for identical cache size problem. 

\subsection{Identical Cache Size and Identical File Length Case}
For the identical cache size and identical file length problem where $M_{1}=\ldots=M_{K}=M$, $F_{1}=\ldots=F_{K}=1$
the optimal memory-rate tradeoff is denoted by $L^{*}(M)$.
%=L^{*}(M,\ldots,M)$.

%\subsection{Uncoded caching inner bound}\label{sub:Uncoded-caching-inner}
The simplest achievable scheme in this case is: in the placement
phase users cache a copy of a fraction $M/N$
of each file, and in the delivery
phase the server only sends the remaining part of the requested file
to each one. With this
%If $M_{1}=M_{2}=...=M_{K}=M$, the load of this method is 
\begin{align}
K(1-M/N)\geq L^{*}(M),
\label{eq:K(1-M/N)}
\end{align}
where $(1-M/N)$ is referred to as the {\it local caching gain.}

%\subsection{Coded caching inner bound}\label{sub:Coded-caching-inner}
%However t
The coded caching strategy of \cite{dvbt2fundamental}
can achieve an additional {\it global caching gain} as follows.
Let %$M_{1}=\ldots=M_{K}=
$M=t\frac{N}{K}$,
for some positive integer $t\in[0:K]$. 
In the placement
phase, each file is split into $B(K,t)$ non-overlapping sub-files
of equal size. The sub-files of $F_{i}$ are denoted by $F_{i,\mathcal{W}}$
for $\mathcal{W}\subseteq[1:K]$ where $|\mathcal{W}|=t$. User $k$ stores $F_{i,\mathcal{W}}$
in his cache if $k\in\mathcal{W}$. In the delivery phase,
for each subset $\mathcal{S}\subseteq[1:K]$ of size $|\mathcal{S}|=t+1$,
the server transmits $\sum_{s\in\mathcal{S}}{\oplus}F_{d_{s},\mathcal{S}\backslash  s}$.
Note that user $i\in\mathcal{S}$ wants $F_{d_{i},\mathcal{S}\backslash  i}$
and knows $F_{d_{s},\mathcal{S}\backslash  s}$ 
%where $s\in\mathcal{S}$ and 
for all $s\neq i$, so he can recover $F_{d_{i},\mathcal{S}\backslash  i}$.
%. So receiving $\sum_{s\in\mathcal{S}}{\oplus}F_{d_{s},\mathcal{S}\backslash  s}$,
%user $i$  can restore $F_{d_{i},\mathcal{S}\backslash  i}$. 
As a result,
%if the server transmits $\sum_{s\in\mathcal{S}}{\oplus}F_{d_{s},\mathcal{S}\backslash  s}$
%for all $\mathcal{S}$ whose length is $t+1$, each user can get all
%his desired sub-files. The 
the load %of this method is equal to
satisfies
\begin{align}
K(1-M/N)\frac{1}{1+KM/N} \geq L^{*}(M).
\label{eq:innerboundofcodedcaching}
\end{align}
Comparing~\eqref{eq:innerboundofcodedcaching} with~\eqref{eq:K(1-M/N)}
%the uncoded caching load, 
the additional global caching gain $\frac{1}{1+KM/N}$ is obtained.

When $t\frac{N}{K}<M<(t+1)\frac{N}{K}$,
time-sharing can be used between the two achievable
loads for $M=t\frac{N}{K}$
and $M=(t+1)\frac{N}{K}$. So the load is a piecewise linear curve. 

Considering the natural multicasting opportunity in the cases $N<K$,
the load in~\eqref{eq:innerboundofcodedcaching} can be improved to
\begin{align}
%L_{C}(M):=
K(1-M/N)\cdot\min\left\{ \frac{1}{1+KM/N},\frac{N}{K}\right\} 
\geq L^{*}(M).
\label{eq:LC(M)}
\end{align}

%Note that with fixed $N$, $K$ and $M$ for all the possible demands the loads of this method are the same.

%\subsection{Outer bound}
In \cite{dvbt2fundamental}, the authors derived a cut-set type outer bound as well.
The optimal load $L^{*}(M)$ must satisfy
\begin{align}
L^{*}(M)\geq \max_{s\in[1:\min(N,K)]}
\left(s-\frac{s}{\left\lfloor N/s\right\rfloor }M\right).
\label{eq:cutsetbound}
\end{align}

%%%%%%%%%%%A cut-set outer bound proposed 
%%%%%%%%%%%In \cite{dvbt2fundamental} it was also shown that
%%%%%%%%%%%%claims
%%%%%%%%%%%%that If $M_{1}=...=M_{K}=M$, the optimal load $L^{*}(M)$ should
%%%%%%%%%%%%satisfy
%%%%%%%%%%%\begin{align}
%%%%%%%%%%%L^{*}(M)\geq\sum_{s\in[1:\min\{N,K\}]}\max\left(s-\frac{s}{\left\lfloor N/s\right\rfloor }M\right)
%%%%%%%%%%%\label{eq:L*(M)}
%%%%%%%%%%%\end{align}
%%%%%%%%%%%%\cite{criticaldatabase}, \cite{improvedlower}, \cite{noteonfundamental}
%%%%%%%%%%%%and \cite{ISIT2015outerbound} proposed tighter outer bounds. \cite{improvedlower}
%%%%%%%%%%%%and \cite{noteonfundamental} use two different computational approaches
%%%%%%%%%%%%algorithms while \cite{criticaldatabase} and \cite{ISIT2015outerbound}
%%%%%%%%%%%%consider some special request-cache tuples and then use information
%%%%%%%%%%%%theoretic derivation.

%\subsection{Coded placement phase}
%
%The ideas in \ref{sub:Uncoded-caching-inner} and \ref{sub:Coded-caching-inner}
%are based on uncoded placement phase. In the placement phase, each
%user directly stores some sub-files. When we use coding in the placement
%phase, we can get a tighter inner bound, 
The load in~\eqref{eq:LC(M)} with {\it uncoded cache placement} can be improved
for instance by storing linear combination of sub-files.
In \cite{smallbufferusers} it was shown that when $K\geq N$ and $M\leq\frac{1}{K}$
the outer bound in~\eqref{eq:cutsetbound} can be achieved as follows.
When $M=\frac{1}{K}$, each file $j$ is split
into $K$ disjoint parts $(F_{j,1},F_{j,2},\ldots,F_{j,K})$. User $i$
stores $\sum_{j\in[1:N]}{\oplus}F_{j,i}$ in his cache. In the
delivery phase, the server transmits $F_{d_{i},s}$ for each $i\in[1:K]$
and each $s\in[1:K]\backslash  {i}$. The achievable load is
%\begin{align}
%L^{*}\left(\frac{1}{K}\right) = N(1-\frac{1}{K}).
%\label{eq:N(1-N/K)}
%\end{align}
%
%
%When $0<M<\frac{1}{K}$, time-sharing is used such that the memory-rate
\begin{align}
L^{*}(M)=N(1-M), \quad KM\leq1, \ K\geq N.
\label{eq:L*M=00003DN(1-M)}
\end{align}

\subsection{Connection to Index Coding}
Consider uncoded cache placement.
When the cache contents and the users' demands are given, 
the caching problem becomes an index coding problem. 
Each file is divided into sub-files and each sub-file is demanded by a new user
% can be seen as an independent user in the index coding system. This new user demands the related sub-file and 
who has the same side information as the original user who demands this sub-file. 
%The main contribution in this report
%is to utilize the index coding results into the caching problem. So
%in the following we will introduce the index coding problem proposed
%in \cite{onthecapacityindex}.
Therefore, known outer bounds for the index coding problem can be used to study the ultimate performance of uncoded cache placement.

In the index coding problem a sender wishes to communicate an independent message $M_{j}, \ j\in[1:N],$ uniformly distributed over $[1:2^{n R_{j}}]$, to the $j$-th receiver  by
broadcasting a message $X^{n}$ of length $n$. 
Each receiver $j$ knows a set of messages, indicated as $\mathcal{A}_j$.
A rate vector $(R_{1},\ldots,R_{N})$ is achievable, for large enough $n$,
if every user can restore his desired message
with high possibility based on $X^{n}$ and his side information.
The index coding problem can be represented as a directed
graph $G$: each node in the graph represents one user; 
a directed edge connects $i$ to $j$ if user $j$ knows $M_{i}$.

A cut-set-type outer bound from \cite{onthecapacityindex} is:
%based on the directed graph of index coding problem.
\begin{thm}[\cite{onthecapacityindex}]%[{\cite[Corollary 1]{onthecapacityindex}]
\label{thm1 uncycle bound} 
If $(R_{1},\ldots,R_{N})$ is achievable
for the index coding problem represented by the directed graph $G$,
then it must satisfy
\begin{align}
\sum_{j\in\mathcal{J}} \frac{|M_{j}|}{n}
=
\sum_{j\in\mathcal{J}} R_{j} \leq 1
\label{eq:Rj<1}
\end{align}
for all $\mathcal{J}\subseteq[1:N]$ where the sub-graph of $G$ over
$\mathcal{J}$ does not contain a directed cycle.
Here $|M_{j}|$ indicates the length in bits of the message for receiver $j$.
\end{thm}

%The rate constraint in $~\eqref{eq:Rj<1})$ can also be expressed as
%\begin{align}
%\sum_{j\in\mathcal{J}} nR_{j}\leq n\label{eq:nRj<n}
%\end{align}
%where $nR_{j}$ represents the number of bits of message $M_{j}$
%and $n$ represents the number of broadcast bits.
%\end{rem}
%$~\eqref{eq:Rj<1})$ can also expressed as 
%\begin{align}
%\sum_{j\in\mathcal{J}} nR_{j}\leq n\label{eq:nRj<n}
%\end{align}
%where $nR_{j}$ represents the number of bits of message $M_{j}$
%and $n$ represents the number of broadcast bits.

\section{Main Result}
%Proof the optimality the scheme in in general case under the constraint of uncoded placement}
\label{sec:main}
Our main results are as follows.

\begin{thm}
For the setting of \cite{dvbt2fundamental},
%the identical cache size and identical file length problem, if $\frac{N(K-1)}{K}\leq M\leq N$, it is optimal to use the coded caching scheme in \cite{dvbt2fundamental} with
the load in~\eqref{eq:innerboundofcodedcaching} attains the outer bound in~\eqref{eq:cutsetbound} when $\frac{N(K-1)}{K}\leq M\leq N$.
\end{thm}
\begin{IEEEproof}
From the cut-set bound in~\eqref{eq:cutsetbound} with $s=1$ we have 
\begin{align}
L^{*}(M)\geq1-\frac{M}{N}.
\label{eq:1-m/n}
\end{align}
The bound in~\eqref{eq:1-m/n} contains the points $(M,L)=(\frac{N(K-1)}{K},\frac{1}{K})$ and $(M,L)=(N,0)$.  The point $(M,L)=(N,0)$ is trivially achievable (each cache can store all files).
When $M=\frac{N(K-1)}{K}$
%, the coded caching in \cite{dvbt2fundamental} gives us 
the load in~\eqref{eq:innerboundofcodedcaching} equals
$\frac{1}{K}$.  Thus the scheme of \cite{dvbt2fundamental}
is optimal for $\frac{N(K-1)}{K}\leq M\leq N$.
\end{IEEEproof}

\begin{thm}%[{\cite[Corollary 1]{onthecapacityindex}]
\label{thm:main} 
The minimal load of the worst case among all the possible demands under the constraint of uncoded cache placement and $N\geq K$ for the case where the total file size and the total cache size are fixed, 
is achieved by letting each user have the same cache size and each file have the same length, then using the coded caching in \cite{dvbt2fundamental} with the load in~\eqref{eq:innerboundofcodedcaching}.\end{thm}

The rest of the section is devoted to the proof of Theorem \ref{thm:main}.
We note that for the case $N<K$ a sub-file may be demanded by more than one users;
so the messages represented by the nodes in the index coding graph are not independent;
as a result, Theorem \ref{thm1 uncycle bound} from \cite[Corollary 1]{onthecapacityindex} can not be used.
Before we present the actual proof, we give an example to highlight the key ideas behind this proof.

\subsection{Example for $N=K=3$}
\label{sub:example N=K=3}

\begin{figure}
\centering
\includegraphics[scale=0.4]{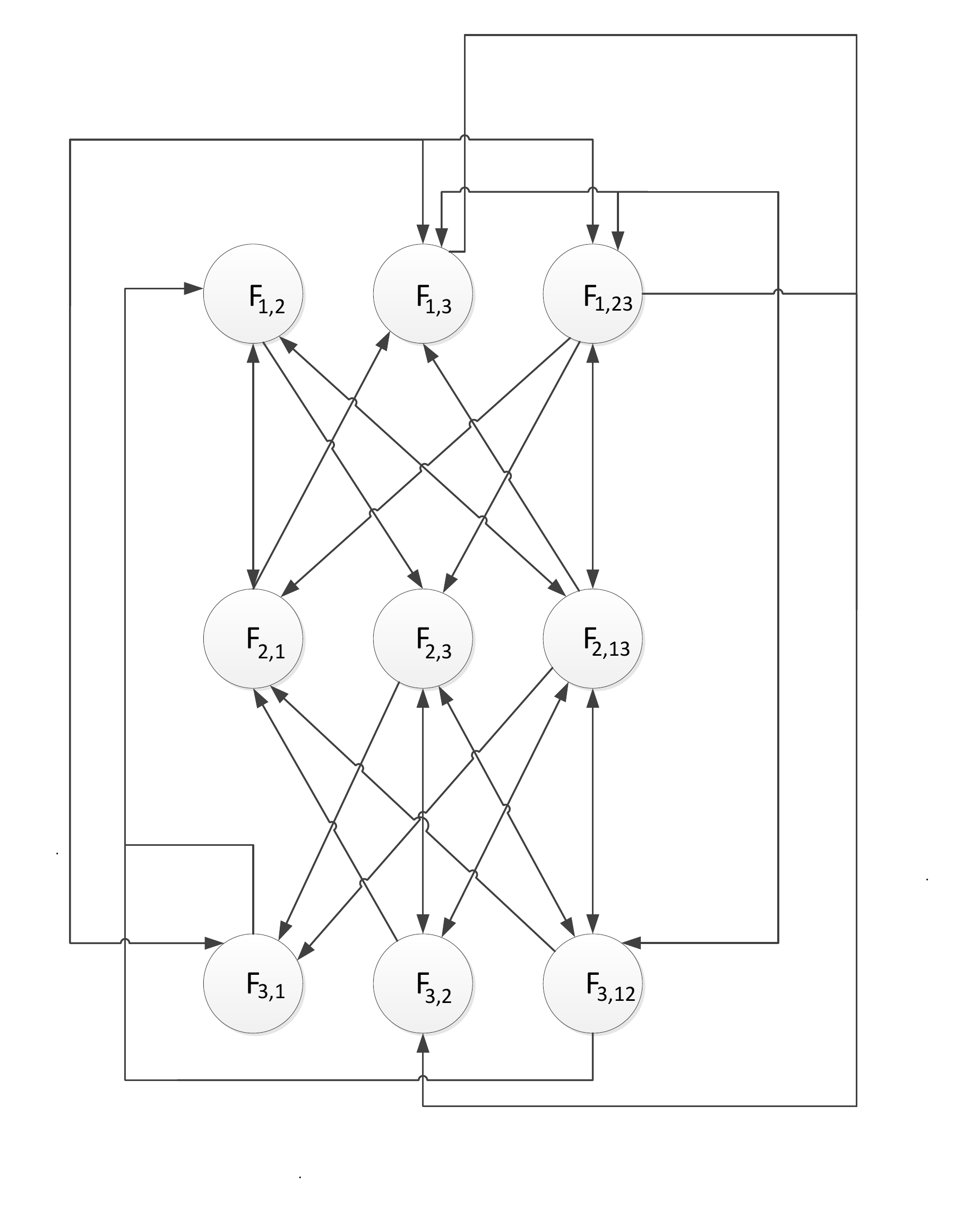}%\protect
\caption{Directed graph for for the equivalent index coding problem in Example~1 with demand vector $(1,2,3)$.}
\label{fig:case3graph}
\end{figure}

%In the following we will show how to derive $~\eqref{eq:after using lemm1})$ in an example.
%\begin{example}
Assume that the server has $N=K=3$ files $(F_{1},F_{2},F_{3})$.
The total file length is $\sum_{i\in[1:3]}|F_{i}| \geq N=3$.
The total cache size is $\sum_{i\in[1:3]}|M_{i}| \leq KM=3M$, 
for some $ M\in[0,N]=[0,3]$.
Each file $F_{i}$ is divided into $2^{K}=2^3=8$ disjoint parts, 
denoted as $F_{i,\mathcal{W}}, \ \mathcal{W} \in 2^{[3]}$
where $2^{[3]}$ indicates the power set 
$2^{[3]} = \{\emptyset, \{1\},\{2\},\{3\}, \{1,2\},\{1,3\},\{2,3\}, \{1,2,3\}\}$. $F_{i,\mathcal{W}}$ is only known by the users in $\mathcal{W}$. For simplicity in the following we omit the braces when we indicate sets, i.e., $F_{1,12}$ represents $F_{1,{\{1,2\}}}$.

Consider the demand vector $\mathbf{d}=(d_{1},d_{2},d_{3})\in[1:3]^3$, where $d_{i}\neq d_{j}$ for all $i\neq j$. %and $d_{1},d_{2},d_{3}\in[1:K]$. 
According to $\mathbf{d}$, user $1$, $2$, $3$ require $F_{d_{1}}$, $F_{d_{2}}$, $F_{d_{3}}$, respectively.  
For each demand vector $\mathbf{d}$
%=(d_{1},d_{2},\ldots,d_{K})$ where $d_{i}\neq d_{j}$ for all $i\neq j$, we can 
we generate an index coding problem with $K 2^{K-1}=12$ independent messages, each of which represents a sub-file demanded by a user in the caching system who does not have it his cache. For this index coding problem, we can generate a directed graph including $12$ nodes as follows. Each node corresponds to one different sub-file/message. We denote the user in the caching system who wants sub-file $j$ by $P_{j}$. There is a directed edge from node $i$ to node $j$ if user $P_{j}$ knows sub-file $i$. 

In order to apply Theorem~\ref{thm1 uncycle bound}, in the constructed graph we want to find the sets that do not containing a cycle. 
Nobody knows $F_{1,\emptyset},F_{2,\emptyset},F_{3,\emptyset}$ so there is
no outgoing edge from $F_{1,\emptyset},F_{2,\emptyset},F_{3,\emptyset}$ to any other nodes. Therefore, $F_{1,\emptyset},F_{2,\emptyset},F_{3,\emptyset}$ are always in the such sets $\mathcal{J}$ when we evaluate~\eqref{eq:Rj<1}. 
For clarity of representation, we do not draw $F_{1,\emptyset},F_{2,\emptyset},F_{3,\emptyset}$ in the directed graph representing the index coding problem.
In Fig.~\ref{fig:case3graph} we draw such a graph for $\mathbf{d}=(1,2,3)$.

%In this figure, we define $\mathbf{u}=(u_{1},u_{2},u_{3})$ where $u_{i}\neq u_{j}$ for all $i\neq j$ and $u_{1},u_{2},u_{3}\in[1:K]$. 
For a demand vector $\mathbf{d}$, consider now permutations $\mathbf{u}=(u_{1},u_{2},u_{3})$ of $\{1,2,3\}$.
For each $\mathbf{u}$, a set of nodes not containing a cycle is as follows: 
$F_{d_{u_{1}},\mathcal{W}_{1}}$ for all $\mathcal{W}_{1}\subseteq[1:3]\backslash \{u_{1}\}$, and
$F_{d_{u_{2}},\mathcal{W}_{2}}$ for all $\mathcal{W}_{2}\subseteq[1:3]\backslash \{u_{1},u_{2}\}$, and 
$F_{d_{u_{3}},\mathcal{W}_{3}}$ for all $\mathcal{W}_{3}\subseteq[1:3]\backslash \{u_{1},u_{2},u_{3}\}=\emptyset$. 
For example, when $\mathbf{d}=(1,2,3)$ and $\mathbf{u}=(1,3,2)$, 
\begin{align*}
&d_{u_1}=d_{1}=1; \mathcal{W}_1\subseteq[1:3]\backslash \{u_1\}=[1:3]\backslash \{3\}=\{2,3\},
\\
&%j=2: 
d_{u_2}=d_{3}=3; \mathcal{W}_2\subseteq[1:3]\backslash \{u_1,u_2\}=[1:3]\backslash \{3,1\}=\{2\},
\\
&%j=3: 
d_{u_3}=d_{2}=2; \mathcal{W}_3\subseteq[1:3]\backslash \{u_1,u_2,u_3\}%=[1:3]\backslash \{3,1,2\}
=\emptyset,
\end{align*}
the corresponding set not containing a cycle is $(F_{1,\emptyset},F_{1,2},F_{1,3},F_{1,23},F_{3,2},F_{3,\emptyset},F_{2,\emptyset})$, as it can be easily verified by inspection of Fig.~\ref{fig:case3graph};
from \eqref{eq:Rj<1} we have that this set implies the bound
\[
n\geq|F_{1,\emptyset}|+|F_{1,2}|+|F_{1,3}|+|F_{1,23}|+|F_{3,2}|+|F_{3,\emptyset}|+|F_{2,\emptyset}|.
\]

In general, we can find such a bound for all possible pairs $(\mathbf{d},\mathbf{u})\in[3!]^2$ 
(here $[3!]$ denotes the set of all permutations of the integers $[1:3]$;
there are $3!$ elements in the set $[3!]$).

We then sum all the $(3!)^2$ inequalities and get
\begin{align}
n (3!)^2 &\geq
\sum_{\mathbf{d} \in [3!]} %demands
\sum_{\mathbf{u} \in [3!]} %no-cycle
\sum_{j \in [3] } %users
\sum_{\mathcal{W}_j \in 2^{[3]} : \mathcal{W}_j \backslash \{u_1,...,u_j\}}
|F_{d_{u_{j}}, \mathcal{W}_j}|
\nonumber\\&=
(3!)^2 \sum_{i \in [0:3]}
x_{i} \frac{1-i/3}{1+i},
\nonumber\\& {\Longleftrightarrow}
n \geq 1 \cdot x_{0}+\frac{1}{3} \cdot x_{1}+\frac{1}{9} \cdot x_{2}+0 \cdot x_{3},
\label{eq:case3inequal}
\end{align}
where
\begin{align}
0\leq x_{t}:=\sum_{j\in[1:N]} \ \sum_{\mathcal{W}\subseteq[1:K] : |\mathcal{W}|=t}   |F_{j,\mathcal{W}}|, \quad t \in [0:K],
\label{eq:defxi}
\end{align}
is the total length of the sub-files that are known by subsets of $t$ users,
and where the proof of why the $x_{t}$'s are multiplied by $\frac{1-t/3}{1+t}$ will be given in the next sub-section.

We also have the sum file size constraint
\begin{align}
&
3 \leq  \sum_{j\in[1:3]} \sum_{\mathcal{W}\subseteq[1:3]}  |F_{j,\mathcal{W}}|
\nonumber\\&
{\Longleftrightarrow}
3 \leq x_0+x_1+x_2+x_3,
\label{eq:casefilesize}
\end{align}
and the sum cache size constraint
\begin{align}
&
\sum_{j\in[1:3]} \sum_{\mathcal{W}\subseteq[1:3] : j\in \mathcal{W}}  |F_{j,\mathcal{W}}| \leq 3M
\nonumber\\&
{\Longleftrightarrow}
0\cdot x_0+1\cdot x_1+2\cdot x_2+3\cdot x_3 \leq 3M.
\label{eq:case3cachesize}
\end{align}

The bounds in~\eqref{eq:case3inequal}-\eqref{eq:case3cachesize}
%,~\eqref{eq:casefilesize}, and~\eqref{eq:case3cachesize} 
provide an outer bound for the load $n$ with uncoded cache placement. 
Since we have many bounds/inequalities in four unknowns, we proceed to 
%do Fourier-Motzkin
eliminate $(x_0,x_1,x_2,x_3)$ in the system of inequalities in~\eqref{eq:case3inequal}-\eqref{eq:case3cachesize}. 
By doing so, we obtain
\begin{align}
n &\stackrel{\text{eq.\eqref{eq:case3inequal}}}{\geq} 
x_{0}+\frac{1}{3}  x_{1}+\frac{1}{9} x_{2}
\nonumber\\&
   \stackrel{\text{eq.\eqref{eq:casefilesize}}}{\geq}
    (3-x_1-x_2-x_3)+\frac{1}{3}  x_{1}+\frac{1}{9} x_{2}
\nonumber\\&
   = 3 - \frac{2}{3} x_{1}-\frac{8}{9} x_{2}-x_3
\nonumber\\&
  \stackrel{\text{eq.\eqref{eq:case3cachesize}}}{\geq}
   3 + \frac{2}{3}(2x_2+3x_3-3M)-\frac{8}{9} x_{2}-x_3
\nonumber\\&
   = 3-2M + \frac{2}{9} x_2 + x_3
   \geq 3-2M,
 \label{eq:case3eq1}
\end{align}
and similarly
\begin{align}
n &\geq  -\frac{2}{3}M+\frac{5}{3},
 \label{eq:case3eq2}
 \\
n  &\geq  -\frac{1}{3}M+1.
 \label{eq:case3eq3}
\end{align}

The maximum among the right-hand sides of~\eqref{eq:case3eq1},
~\eqref{eq:case3eq2}
and~\eqref{eq:case3eq3}
give a piecewise linear curve with corner points
$(0,3),(1,1),(2,\frac{1}{3}),(3,0)$.
Since these corner points are achieved by~\eqref{eq:innerboundofcodedcaching} and for all possible demand vectors the loads of this scheme are the same, it is optimal for the example problem to firstly let each user have the same cache size and let each file have the same length, then to use the two-phase strategy in \cite{dvbt2fundamental}.

\subsection{General Proof of Theorem 3}
\label{sub:N=K}
The general case $N\geq K$ is proved by a similar method as in the previous example. 
Firstly we consider the case where the file demanded by each user
is different. Note that for this case there are $P(N,K)$ demand vectors, each of which is $\mathbf{d}=(d_{1},d_{2},\ldots,d_{K})$ where $d_{i}\in [1:N]$ and $d_{i}\neq d_{j}$ for all $i\neq j$.  
%We assume that the total cache size of all users is equal to $KM$.
We divide file $F_{j}$ into $2^{K}$ disjoint parts, each of which is
denoted by $F_{j,\mathcal{W}}$ such that
%. Obviously, $0\leq|F_{j,\mathcal{W}}|\leq1$ and
$\sum_{\mathcal{W}\subseteq[1:K]}|F_{j,\mathcal{W}}|=|F_{j}|$.
%where $|F_{j,\mathcal{W}}|$ means the length of $F_{j,\mathcal{W}}$.
$F_{j,\mathcal{W}}$ is only known by the users in $\mathcal{W}$.
For each demand vector, we generate a directed graph
with $K 2^{K-1}$ nodes as the same method claimed in the previous example.

We construct cycles in the directed graph by the following Lemma.
\begin{lem}
\label{lem:For-each-graph}
Let $\mathbf{u}=(u_{1},u_{2},\ldots,u_{K})$ be a permutation of $[1:K]$.
%where $u_{i}\neq u_{j}$ for all $i\neq j$ and $u_{i}\in[1:K]$. 
%So for each $\mathbf{u}$, 
A set of nodes not containing a cycle in the directed graph of the corresponding
index coding problem contains sub-file
$F_{d_{u_{i}},\mathcal{W}_{i}}$ for all $i\in[1:K]$ and all $\mathcal{W}_{i}\subseteq[1:K]\backslash \{u_{1},\ldots,u_{i}\}$.
%
%$F_{d_{u_{1}},\mathcal{W}_{u_{1}}}$ for all $\mathcal{W}_{u_{1}}\subseteq[1:K]\backslash \{u_{1}\}$,
%$F_{d_{u_{2}},\mathcal{W}_{u_{2}}}$ for all $\mathcal{W}_{u_{2}}\subseteq[1:K]\backslash \{u_{1},u_{2}\}$,\ldots,
%$F_{d_{u_{K}},\emptyset}$. 
\end{lem}
\begin{IEEEproof}
For a $\mathbf{u}=(u_{1},u_{2},\ldots,u_{K})$, we say that sub-files/nodes $F_{d_{u_{i}},\mathcal{W}_{i}}$ for all $\mathcal{W}_{i}\subseteq[1:K]\backslash \{u_{1},...,u_{i}\}$ are in level $i$. It is easy to see each node in level $i$ only knows the sub-files $F_{j,\mathcal{W}}$ where $u_{i}\in\mathcal{W}$. So each node in level $i$ knows neither the sub-files in the same level, nor the sub-files in the higher levels. As a result, in the proposed set there is no sub-set containing a directed cycle. 
\end{IEEEproof}

According to~\eqref{eq:Rj<1} and Lemma \ref{lem:For-each-graph},
in order to recover all the desired sub-files for each user, the number
of broadcast bits $n$ needs to satisfy
\begin{align}
n & \geq\sum_{\mathcal{W}_{1}\subseteq[1:K]\backslash \{u_{1}\}}|F_{d_{u_{1}},\mathcal{W}_{1}}|+...+\sum_{\mathcal{W}_{i}\subseteq[1:K]\backslash \{u_{1},\ldots,u_{i}\}}\nonumber \\
 & |F_{d_{u_{i}},\mathcal{W}_{i}}|+...+\sum_{\mathcal{W}_{K}\subseteq[1:K]\backslash \{u_{1},\ldots,u_{K}\}}|F_{d_{u_{K}},\mathcal{W}_{K}}|.
 \label{eq:original uncycle}
\end{align}
Considering all the possible demands vectors where $d_{i}\neq d_{j}$ if $i\neq j$
and all the $\mathbf{u}$ for each demand vector, we can list all the inequalities
in the form of~\eqref{eq:original uncycle}. 
%There are $N!$ demand vectors and for each demand there are $N!$ vectors $\mathbf{u}$. 
%So the number of considered 
There are $P(N,K)P(K,K)$ such inequalities. 
Because of symmetry,
for each $i\in[0:K]$ on the right side of the sum of all the $P(N,K)P(K,K)$ inequalities
the coefficients of the term $|F_{j,\mathcal{W}}|$, for $j\in[1:N]$ and $|\mathcal{W}|=i$,
are equal. 

In~\eqref{eq:original uncycle} there are $B(K-1,i)+B(K-2,i)+...+B(i,i)$
terms with $|\mathcal{W}|=i$ whose coefficient is $1$.
Since there are totally $B(K,i)N$ sub-files with $|\mathcal{W}|=i$,
in the sum expression the coefficient of each $|F_{j,\mathcal{W}}|$
with $|\mathcal{W}|=i$ is $P(N,K)P(K,K) \frac{B(K-1,i)+...+B(i,i)}{B(K,i)N}$.
As a result we have
\begin{align}
n\geq \sum\limits_{i=0}^K \frac{B(K-1,i)+...+B(i,i)}{B(K,i)N}x_{i}.
\label{eq:pn>}
\end{align}
From the Pascal's triangle %, $B(N,i)=B(N+1,i+1)-B(N,i+1)$,
we have
\begin{align}
B(K-1,i)+...+B(i,i)=B(K,i+1),
\label{eq:using lemma1}
\end{align}
thus we rewrite~\eqref{eq:pn>} as
\begin{eqnarray}
n &\geq  \sum\limits_{i=0}^K \frac{B(K,i+1)}{B(K,i)N}x_{i}
=  \sum\limits_{i=0}^K \frac{K-i}{(i+1)N}x_{i}
\label{eq:after using lemm1}
\end{eqnarray}
Since the total size of all files is  
\begin{align}
x_{0}+x_{1}+...+x_{K} \geq N
\label{eq:totalsizeN}
\end{align}
and the total cache size is
\begin{align}
x_{1}+2x_{2}+...+ix_{i}+...+Kx_{K} \leq KM,
\label{eq:total cache NM}
\end{align}
we obtained the desired bound on $n$.

In Appendix, we prove the by combining the derived bound we can write:
%\label{lem:foreachq}
for each $q\in[1:K]$, 
\begin{align}
&n\geq\frac{-(K+1)KM}{Nq(q+1)}+\frac{2K-q+1}{q+1}+\sum\limits_{i=0}^K Z(N,K,i,q)x_{i}
\label{eq:TH2}
\\
&Z(N,K,i,q)=\frac{(K+1)(i-q+1)(i-q)}{qN(q+1)(i+1)}.
\label{eq:TH2next}
\end{align}

Note that $x_{i}$ depends on $M$ and for each $M$ we need not strictly
find the maximum of the right side of~\eqref{eq:TH2} among all
$q$. Instead, for each $M$ we prove that the right side is achievable
with a $q\in[1:K]$. As a result, this load is the minimum for such
$M$.
From~\eqref{eq:TH2}, for one $q\in[1:K]$ the outer
bound in~\eqref{eq:TH2} becomes linear in terms of $M$. We focus
our attention on $\frac{(q-1)N}{K}\leq M\leq \frac{qN}{K}$.

For $M=\frac{(q-1)N}{K}$ we have
\begin{align*}
n &\geq \frac{-(K+1)KM}{Nq(q+1)}+\frac{2K-q+1}{q+1} =\frac{K-q+1}{q}
\\
 & =K(1-\frac{M}{N})\frac{1}{1+KM/N},
\end{align*}
and for $M=\frac{qN}{K}$ we have
\begin{align*}
n &\geq \frac{-(K+1)KM}{Nq(q+1)}+\frac{2K-q+1}{q+1} =\frac{K-q}{q+1}
\\
 & =K(1-\frac{M}{N})\frac{1}{1+KM/N}.
\end{align*}
$K(1-\frac{M}{N})\frac{1}{1+KM/N}$ is the load achieved by giving
each user the same cache size, giving each file the same length, and 
using the two-phase coded caching in \cite{dvbt2fundamental}. Note that the loads of this scheme for all the possible demand vectors are the same. So we conclude that when $N\geq K$ in order to minimize the worst-case load,
the scheme in \cite{dvbt2fundamental}
is optimal among all schemes with uncoded cache placement.

%If $q-1<M<q$, the load achieved by time-sharing of the two points
%$M=q-1$ and $M=q$ coincides with the linear outer bound with such
%$q$. So for all $0\leq M\leq N$, we can find one $q\in[1:N]$ with
%which the outer bound in~\eqref{eq:TH2}) coincides with the load achieved
%by the coded caching in \cite{dvbt2fundamental}.
%
%We have proved under constraint of uncoded placement considering all
%the demands $\mathbf{d}=(d_{1},d_{2},\ldots,d_{K})$ where $d_{i}\neq d_{j}$
%if $i\neq j$, the identical allocation of cache size and file length with the coded caching in \cite{dvbt2fundamental} is optimal. For all the possible methods, the load of the worst case
%is larger than or equal to the load of the aforementioned case. Moreover
%for all the demands the loads of the coded caching in \cite{dvbt2fundamental}
%are the same. As a result, considering the worst case it is also optimal.
%

\section{Conclusion and Further Work}
\label{sec:conc}

In this paper, we considered the cache problem 
%which covers the identical cache size and identical file length caching problem in
of \cite{dvbt2fundamental}, but where we assumed a total cache size constraint and 
a total file size constraint. By leveraging an outer bound for the index coding 
problem, we proved that under the constraint of uncoded cache placement
and $N\geq K$, for minimizing the worst-case load it is optimal to let each user have the same cache
size and each file have the same length, then to use the coded caching proposed in \cite{dvbt2fundamental}.
Our results show that the only way to improve on the load of \cite{dvbt2fundamental} is by 
coded cache placement.
%We also prove for the identical cache size and identical file length problem the general optimality of the coded caching proposed
%in \cite{dvbt2fundamental} when the cache size is large enough. 

Further work will be in two directions:
study the case where $N<K$
and study achievable loads for
coded cache placement.

\section*{Acknowledgments}
The work of K. Wan and D. Tuninetti is supported by Labex DigiCosme and  in part by NSF~1527059, respectively. 

\appendix
\begin{IEEEproof}
For each $q\in[1:K]$, we want to eliminate $x_{q}$ and $x_{q-1}$ in~\eqref{eq:after using lemm1} by the help of~\eqref{eq:totalsizeN} and~\eqref{eq:total cache NM}.

From~\eqref{eq:totalsizeN}, we have
\begin{align}
\frac{2K-q+1}{N(q+1)}(x_{q-1}+x_{q})\geq \frac{2K-q+1}{N(q+1)}(N-\sum_{i\in[0:K]:i\neq q-1,q} x_{i})
\label{eq:generalstep1}
\end{align}

From~\eqref{eq:total cache NM}, we have
\begin{align}
 & \frac{-(K+1)}{Nq(q+1)}(q-1)x_{q-1}-\frac{K+1}{Nq(q+1)}qx_{q}\nonumber \\
 & \geq \frac{-(K+1)}{Nq(q+1)}KM+\frac{K+1}{Nq(q+1)}\sum_{i\in[0:K]:i\neq q-1,q} ix_{i}
 \label{eq:generalstep2}
\end{align}

Then we sum~\eqref{eq:generalstep1} and~\eqref{eq:generalstep2}, 
\begin{align}
 & \frac{K-q+1}{Nq}x_{q-1}+\frac{K-q}{N(q+1)}x_{q}\nonumber \\
 & \geq \frac{2K-q+1}{N(q+1)}(N-\sum_{i\in[0:K]:i\neq q-1,q} x_{i})+\nonumber \\
 & \frac{-(K+1)}{Nq(q+1)}KM+\frac{K+1}{Nq(q+1)}\sum_{i\in[0:K]:i\neq q-1,q} ix_{i}\nonumber \\
 & =\frac{-(K+1)KM}{Nq(q+1)}+\frac{2K-q+1}{q+1}+\nonumber \\
 & \sum_{i\in[0:K]:i\neq q-1,q} (-\frac{2K-q+1}{N(q+1)}+\frac{(K+1)i}{Nq(q+1)})x_{i}\label{eq:generalstep3}
\end{align}

At last we take~\eqref{eq:generalstep3} into~\eqref{eq:totalsizeN}, we have
\begin{align}
n & \geq\sum\limits_{i=0}^K \frac{K-i}{(i+1)N}x_{i}\nonumber \\
 & =\frac{K-q+1}{Nq}x_{q-1}+\frac{K-q}{N(q+1)}x_{q}\nonumber \\
 & +\sum_{i\in[0:K]:i\neq q-1,q} \frac{K-i}{(i+1)N}x_{i}\nonumber \\
 & \geq \frac{-(K+1)KM}{Nq(q+1)}+\frac{2K-q+1}{q+1}+\sum_{i\in[0:K]:i\neq q-1,q} \nonumber \\
 & (-\frac{2K-q+1}{N(q+1)}+\frac{(K+1)i}{Nq(q+1)}+\frac{K-i}{(i+1)N})x_{i}\nonumber \\
 & =\frac{-(K+1)KM}{Nq(q+1)}+\frac{2K-q+1}{q+1}+\sum\limits_{i=0}^K Z(N,K,i,q)x_{i}\label{eq:finalsetp}
\end{align}
where $Z(N,K,i,q)=\frac{(K+1)(i-q+1)(i-q)}{qN(q+1)(i+1)}$. 
This  concludes the proof.
%So we can see~\eqref{eq:finalsetp}) coincides with Lemma \ref{lem:foreachq}
\end{IEEEproof}

\bibliographystyle{IEEEtran}
\bibliography{IEEEabrv,IEEEexample}

\begin{IEEEbiography}[{\fbox{\begin{minipage}[t][1.25in]{1in}%
Replace this box by an image with a width of 1\,in and a height of
1.25\,in!%
\end{minipage}}}]{Your Name}
 All about you and the what your interests are.
\end{IEEEbiography}

%\begin{IEEEbiographynophoto}{Coauthor}
%Same again for the co-author, but without photo\end{IEEEbiographynophoto}

\end{document}